\documentclass[aps,prstab,twocolumn,superscriptaddress]{revtex4-1}
\usepackage{epsfig}
\usepackage{graphics}
\def\blfootnote{\xdef\@thefnmark{}\@footnotetext} 

\begin{document}


\title{ Kelvin Probe Studies of Cesium Telluride Photocathode for AWA Photoinjector}


\author{Eric Wisniewski}
\affiliation{High Energy Physics Division, Argonne National Laboratory, 9700 S. Cass, Lemont, IL 60439}

\affiliation{Physics Department, Illinois Institute of Technology, 3300 South Federal Street, Chicago, IL 60616}

\author{Daniel Velazquez}
\affiliation{High Energy Physics Division, Argonne National Laboratory, 9700 S. Cass, Lemont, IL 60439}
\affiliation{Physics Department, Illinois Institute of Technology, 3300 South Federal Street, Chicago, IL 60616}

\author{Zikri Yusof}
\email[]{zyusof@anl.gov}
\affiliation{High Energy Physics Division, Argonne National Laboratory, 9700 S. Cass, Lemont, IL 60439}

\author{Linda Spentzouris}
\affiliation{Physics Department, Illinois Institute of Technology, 3300 South Federal Street, Chicago, IL 60616}

\author{Jeff Terry}
\affiliation{Physics Department, Illinois Institute of Technology, 3300 South Federal Street, Chicago, IL 60616}

\author{Katherine Harkay}
\affiliation{Accelerator Science Division, Argonne National Laboratory, 9700 S. Cass, Lemont, IL 60439}



\date{\today}

\begin{abstract}
Cesium telluride is an important photocathode as an electron source for particle accelerators. It has a relatively high quantum efficiency \((>1\%)\), is sufficiently robust in a photoinjector, and has a long lifetime. This photocathode is grown in-house for a new Argonne Wakefield Accelerator (AWA) beamline to produce high charge per bunch (\(\approx\)50 nC) in a long bunch train. Here, we present a study of the work function of cesium telluride photocathode using the Kelvin Probe technique. The study includes an investigation of the correlation between the quantum efficiency and the work function, the effect of photocathode aging, the effect of UV exposure on the work function, and the evolution of the work function during and after photocathode rejuvenation via heating.
\end{abstract}


\maketitle



Cesium telluride ($\mathrm{Cs}_2\mathrm{Te}$) photocathodes are a proven electron source for particle accelerators. They have a high quantum efficiency (10\% at 4.9 eV photon energy), a long lifetime (months) and are robust in a high gradient environment~\cite{Kong1995AIP}.  The new RF photocathode drive gun being commissioned at the Argonne Wakefield Accelerator (AWA) is a high peak-current electron beam source for the new 75 MeV linear electron accelerator, to be used to excite wakefields in dielectric-loaded accelerating (DLA) structures and other novel high-gradient structures~\cite{AWAMOP110}.  A unique requirement of the AWA experimental program is the ability to produce long trains of high-charge bunches, hence the need for a high quantum efficiency (QE) photocathode such as $\mathrm{Cs}_2\mathrm{Te}$. The AWA is producing $\mathrm{Cs}_2\mathrm{Te}$ photocathodes for use in the new high-charge, 1.3 GHz photoinjector~\cite{AWAPAC2011}. In particular, an electron bunch train of 30 bunches with up to 50 nC per bunch is expected to be produced.  The substantial demands on the photocathode necessitate a thorough understanding of the photocathode and its parameters.  The QE at a particular photon energy and the work function ($\phi$) are two important parameters of electron emission.  Here, we present the results of Kelvin probe measurements of the work function on $\mathrm{Cs}_2\mathrm{Te}$ photocathodes. We examined (i) the correlation between the QE and the work function; (ii) how QE and the work function evolved with photocathode aging; (iii) effects of rejuvenation of the photocathode via heating, and (iv) the effects on the work function upon exposure to UV light.

The Kelvin probe method is a non-contact, non-destructive technique that is used to measure the potential difference between a sample and the Kelvin probe tip (reference) when the two are in electrical contact. The tip and the sample are set in a parallel plate capacitor configuration and the circuit is completed through ground connection, thus aligning the Fermi levels of tip and sample. The electrical contact between the tip and the sample causes electron migration from higher- to lower-Fermi-level material, creating an electric field between the tip and sample. The potential associated with this electric field is called the contact potential difference (CPD), which multiplied by the electron charge results in the difference of the work functions of the sample and tip. Hence, knowing the work function of the reference tip and measuring the CPD allows the sample work function to be calculated.  The validity of the work function measurement therefore relies on the calibration of the tip using a known reference.  The theory and details of the method have been described in detail elsewhere~\cite{Surplice1970}.  

\begin{figure}[htp]
 
 \includegraphics[width=0.45\textwidth]{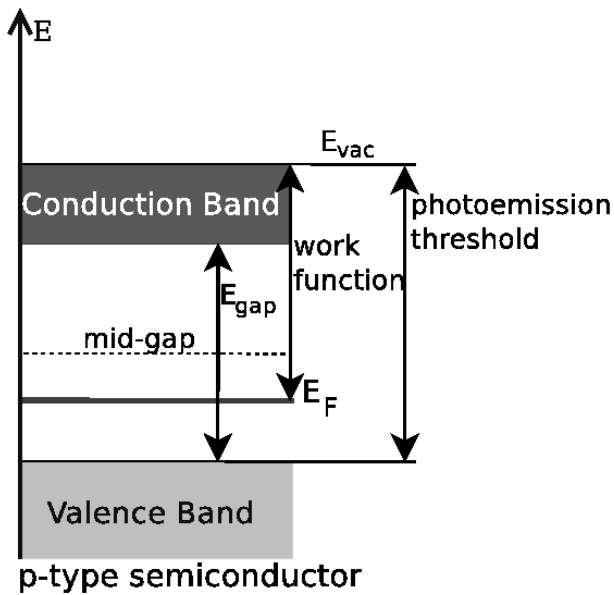}
 \caption{Band diagram for p-type semiconductor.  The work function is measured from the vacuum potential level ($\mathrm{E}_{\mathrm{vac}}$) to the Fermi level ($\mathrm{E}_{\mathrm{F}}$), while the photoemission threshold $(E_t)$ is measured from $\mathrm{E}_{\mathrm{vac}}$ to the valence band maximum.  In this experiment, what is measured is the actual work function and not the photoemission threshold.\label{fig:band_diagram}}
 \end{figure}

Fig.~\ref{fig:band_diagram} shows the band diagram for a p-type semiconductor. The work function is defined as the energy difference between the vacuum potential level and the Fermi level which is located in the energy gap between the valence and conduction bands. On the other hand, the photoemission threshold is defined as the difference between the vacuum level and the valence band maximum. Therefore the work function in a semiconductor is not the same as the photoemission threshold, unlike the case of a metal.  In this experiment, what is measured is the actual work function and not the photoemission threshold.

The photocathodes studied were fabricated in the AWA photocathode laboratory using a standard recipe and procedure~\cite{Kong1995276},\cite{Michelato1997464}.  AWA photocathodes are deposited on a molybdenum plug designed to fit into the back wall of the gun.  In preparation for deposition, the plug is polished and cleaned, placed under vacuum, then heated to $120^{\circ}$C.  A 22 nm layer of tellurium is deposited via thermal evaporation. When tellurium deposition is complete, cesium deposition commences and the photocurrent is monitored. Deposition continues for several minutes after maximum photocurrent is achieved.  The result of this process is a $\mathrm{Cs}_2\mathrm{Te}$ thin film photocathode on a molybdenum substrate with an effective photocathode diameter of 31 mm and a typical initial QE of 15\%.  QE is measured at 4.9 eV photon energy to closely match the photoinjector laser. All QE values reported in this paper were measured using 4.9 eV photon energy.

\begin{figure}[htp]
\includegraphics[width=0.45\textwidth]{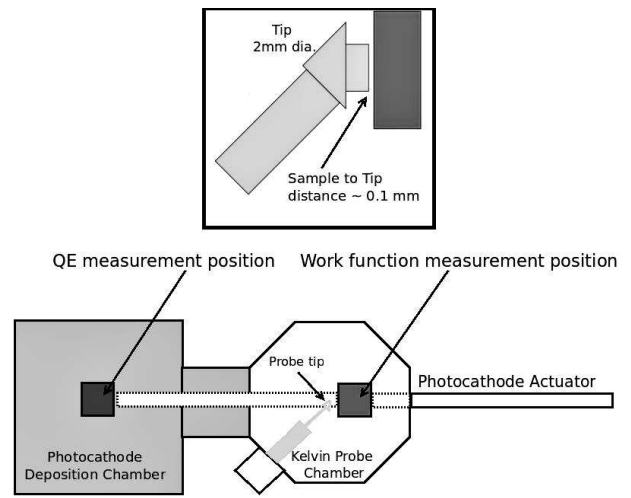}
 \caption{Top view, schematic (not to scale) of experimental setup, showing the Kelvin Probe chamber attached to the back of the deposition chamber.  The actuator is used to move the photocathode from one chamber to the other a distance of \~1.5 m. Inset:  Drawing of the Kelvin probe tip and sample (photocathode) illustrating the relative orientation as seen from above - (zoomed view).\label{fig:ExpSetup}}
 \end{figure}

The experimental setup is pictured in Fig.~\ref{fig:ExpSetup}. It included a large vacuum chamber where the $\mathrm{Cs}_2\mathrm{Te}$ cathodes were fabricated.  The Kelvin probe was housed in the smaller vacuum chamber connected to the back. A long-stroke actuator holding the cathode plug provided the means to easily move the plug back and forth from the deposition chamber (for fabrication and quantum efficiency measurements) to the Kelvin probe chamber (for work function measurements). All QE and Kelvin probe measurements were made in situ. $\mathrm{Cs}_2\mathrm{Te}$ were fabricated and maintained under ultra-high vacuum (UHV) conditions with base pressure of $1.5 \times 10^{-10}$ Torr. 

The Kelvin Probe system is a McAllister Technical Services KP 6500 which includes control software and electronics, and data collection via a PCI National Instruments data acquisition card. The Kelvin probe was positioned in a port in the smaller chamber oriented at $45^{\circ}$ with respect to the sample actuator. In order to keep the surfaces of the KP tip and the sample parallel, the tip was customized to face at $45^{\circ}$ from the longitudinal axis of the tip (see Fig.~\ref{fig:ExpSetup}, inset).  The tip can probe a 2 mm diameter circular area where the work function measurement is the average work function over the probed surface. The coarse sample to tip distance was varied manually using a linear translator attached to the Kelvin probe chamber. The fine adjustment to the sample to tip distance and the tip oscillation along the longitudinal axis of the Kelvin probe were controlled by means of the computer-controlled voice coil system.  The effect of stray capacitances was minimized by doing a spectral analysis to find the resonances of the vibrating probe and subsequently choosing to operate at an off-resonant frequency.


Since the Kelvin probe measures the position of the Fermi level of the sample relative to the reference tip, calibration of the latter is necessary in order to obtain the absolute value of the sample's Fermi level relative to the vacuum level, and thus to be able to obtain the sample's work function. In the setup described here, the tip was made of type 304 stainless steel coated with nichrome, a non-magnetic alloy of nickel and chromium. Calibration was performed using three references of known work function: polycrystalline molybdenum with work function 4.6 eV~\cite{michaelson:4729}, polycrystalline tellurium with work function 4.95 eV~\cite{michaelson:4729},and highly oriented pyrolytic graphite (HOPG) with work function 4.6 eV~\cite{PhysRevB.32.8317},\cite{suzuki:4007}.  In the configuration for calibration the sample played the role of reference and the tip was the material probed, hence the work function of the tip was calculated by adding the contact potential difference measured to the work function of the reference sample.  The work function of the tip was taken as the average of the values found in calibration and the uncertainty taken to be the largest measured. The resulting tip work function value was $4.6 \pm 0.1$~eV.

\begin{figure}[htp]
 
 \includegraphics[width=6cm]{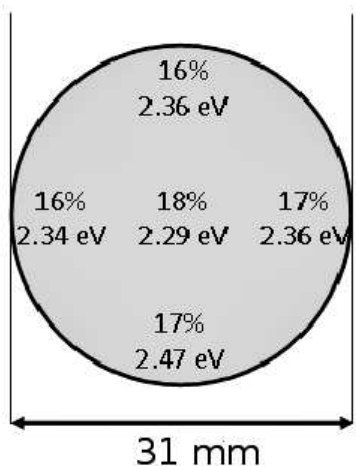}
 \caption{Local variation of photocathode quantum efficiency and work function.  The data were taken within one day of cathode fabrication.\label{fig:cathodemap}}
 \end{figure}

For this experiment, six $\mathrm{Cs}_2\mathrm{Te}$ photocathodes were fabricated and studied. The initial average value of the QE for the cathodes in the study was 16.7\%, the range was [15.5,18.8\%], and standard deviation was 1.3\%. The initial average value of the work function was 2.3 eV, the range was [2.22,2.36], and standard deviation 0.055 eV.   The work function and QE were recorded and tracked for five indexed points on the cathode surface, as shown in Fig.~\ref{fig:cathodemap} for a newly-grown photocathode.  For a photocathode of this size, uniformity in QE could be an issue. As can be seen in Fig.~\ref{fig:cathodemap}, the photocathode that had been fabricated was relatively uniform, both in QE and in the measured work function.  Applying the rigid band picture and using an average measured value of 2.3 eV as the work function for $\mathrm{Cs}_2\mathrm{Te}$, band gap of 3.3 eV and electron affinity of 0.2 eV~\cite{Sommer},\cite{PhysRevB.8.3987} this places the Fermi level $\approx0.5$ eV below the middle of the band gap, making this a p-type semiconductor. 

While it is uncertain if the rigid band model is accurate to describe the
aging effect of QE and the work function, it is still a useful model to
obtain an initial quantitative comparison on how much the Fermi level may
have changed. Thus, applying the same calculation for the typical cathode
after aging for 2-3 weeks, with an average value of the work function of
2.8, the position of the Fermi level is now $\approx1$ eV below the mid-gap, a
shift of 0.5 eV towards the valence band.  Certainly, there may be other
factors that can cause a change in the work function that we measured beyond
just a shift in the Fermi level, including an increase in the electron
affinity due to surface contaminants, etc. We intend to investigate this
further in future studies.

\begin{figure}[htp]
 
  \includegraphics[width=0.45\textwidth]{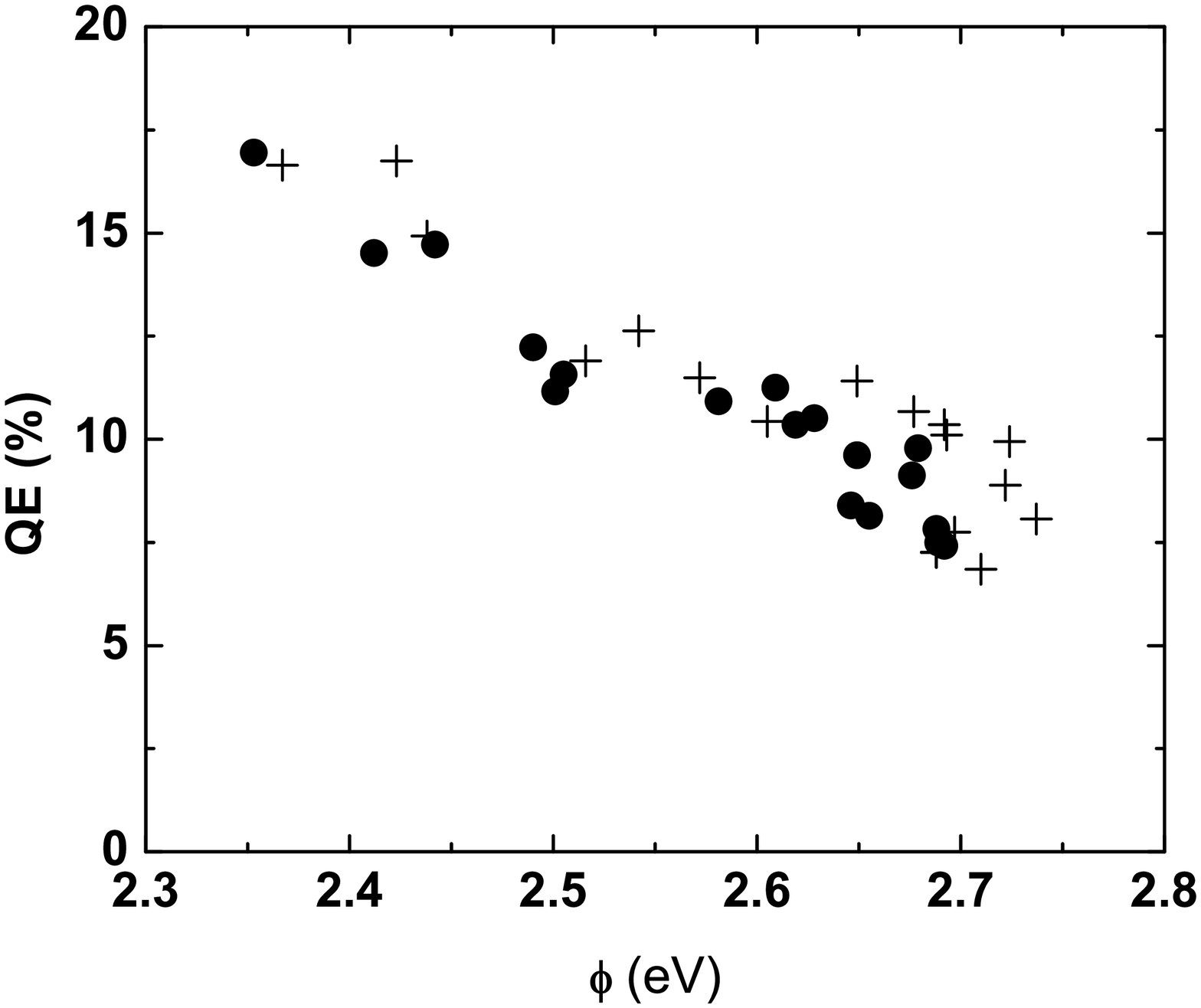}
\hfill
 \includegraphics[width=0.45\textwidth]{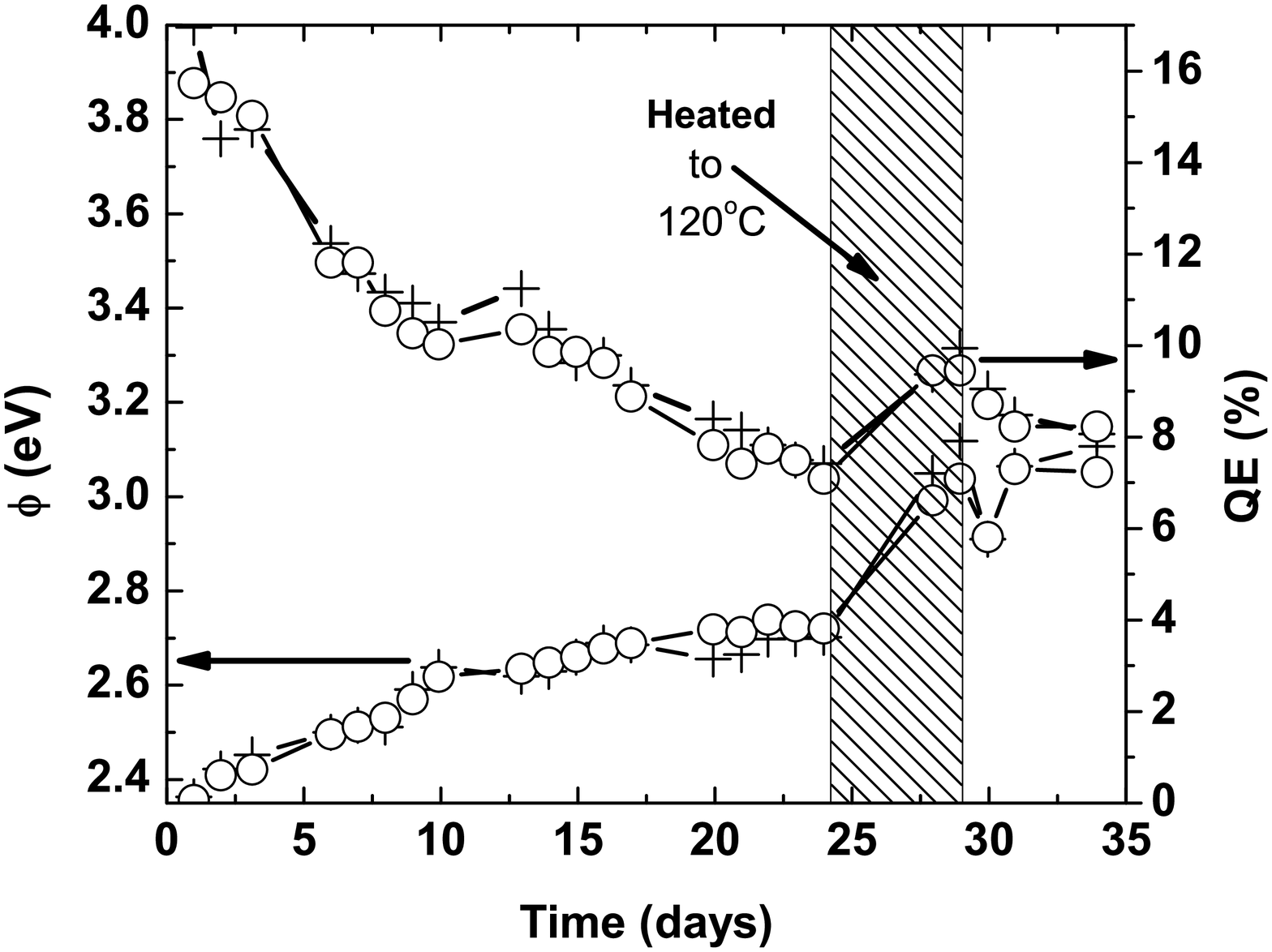}
  \caption{Top: QE vs. work function; typical data. Data taken at two locations on the cathode over a period of about 3 weeks. Bottom: Time evolution of the QE and $\phi$.  The box represents the time period the cathode was heated to $120^{\circ}$C to rejuvenate the QE. Data tracks changes observed at 2 different locations on the cathode.  The work function measurement was performed first followed by QE measurement.  \label{fig:WFvsQE}}
  \end{figure}

It is well-known from previous studies that the QE of $\mathrm{Cs}_2\mathrm{Te}$ diminishes over time~\cite{Kong1995AIP},~\cite{Michelato1997464}.  We investigated the evolution of QE and work function. The plot in Fig.~\ref{fig:WFvsQE}, top shows the correlation between QE and work function at a point on the photocathode over a period of 3 weeks. As the value of the QE dropped, the work function correspondingly increased. The variation of QE and work function over a period of time can be seen in Fig.~\ref{fig:WFvsQE}, bottom. The value for the QE initially dropped rapidly and then started to level off after 15 days. The work function followed this pattern inversely, and appeared to change very little after 20 days.  The observed trend of increased work function with decreasing QE is similar to that observed previously~\cite{SLederer2007} for $\mathrm{Cs}_2\mathrm{Te}$ photocathode after operation. 

A fit of QE vs. work function using a power law for photoemission has been done for metals and semiconductors\cite{Kane1962}, using  $QE=A(E_t-h \nu)^{P}$; where $h \nu$ is the photon energy, $E_t$ is the photoemission threshold, $P$ is a fit parameter ($P$=2 for metals).  Since we probe the work function and not the photoemission threshold, we make the simplification of $\phi=E_t$ in attempting this fit. The result was not very meaningful, yielding values of P in the range of 2.5-4.6, outside of what is expected theoretically. While it is
unambiguously clear that there is an inverse relationship between QE and work function, we are not able to make a quantitative determination of this exact
relationship based on our available data at present. This is something we intend to investigate further in the future.

After aging, the photocathode was rejuvenated via heating at $120^{\circ}$ C for 4 days.\footnote{One photocathode was not a candidate for rejuvenation due to poor vacuum.} Previous studies on photocathode rejuvenation via heating have shown a QE recovery up to about 60\% of the original value~\cite{Kong1995276},~\cite{Bona}.  In our experiment, five of six photocathode's QE went from 30\% of the original QE prior to heating to an average of about 60\% of the original value after heating.  Curiously, however, this increase in QE after rejuvenation was accompanied by an increase in the work function. This was contrary to the pattern seen in Fig.~\ref{fig:WFvsQE}, top, where QE and work function were inversely related during the aging process.

There are many possible explanations for such an observation, including the possibility that the process of heating has changed the chemistry or nature of the photocathode, especially on the surface, resulting in an increase in the photocathode's electron affinity~\cite{Yates2011}. More studies are required to determine the cause of this unexpected behavior.

\begin{figure}[htp]
\includegraphics[width=0.45\textwidth]{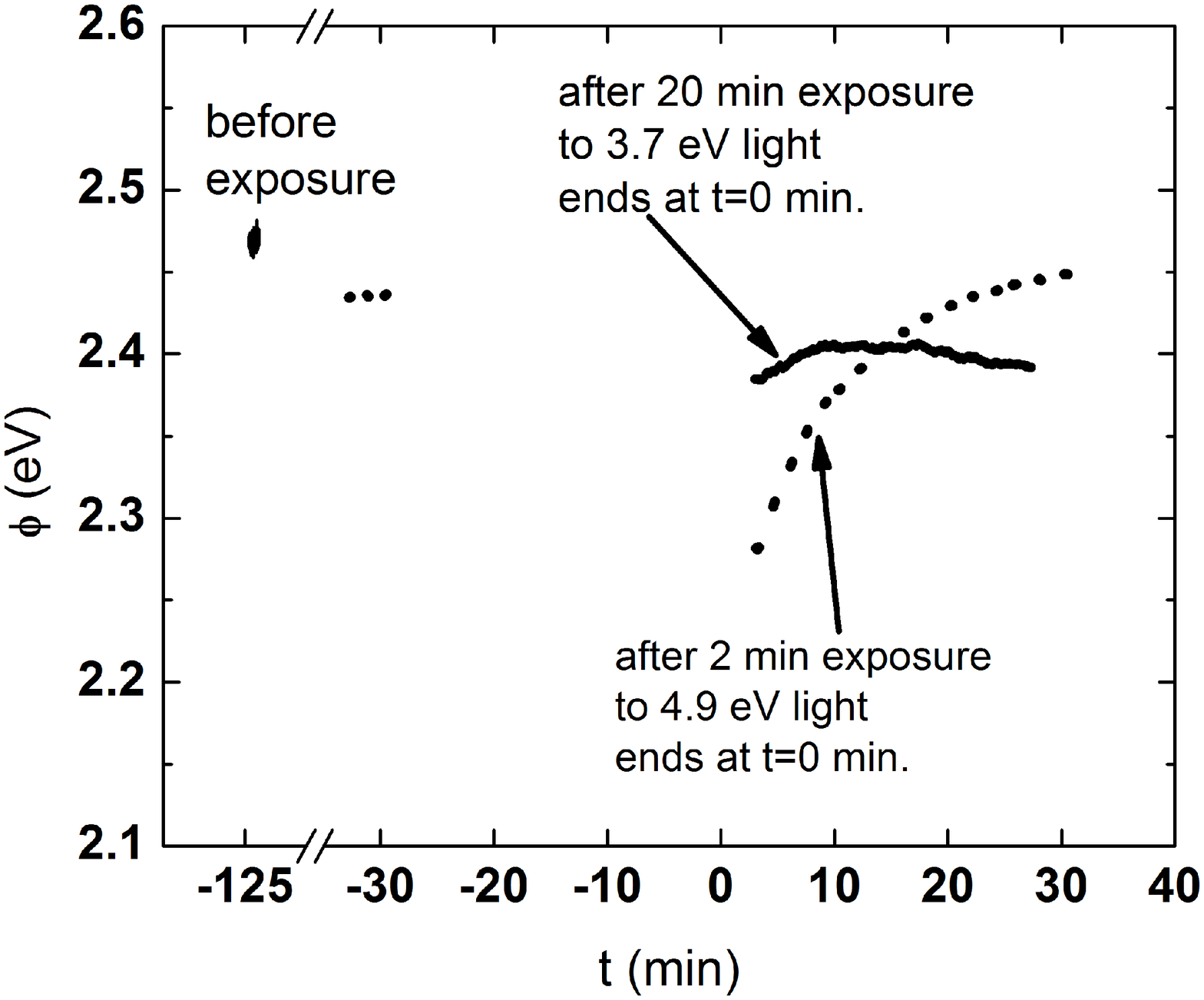}
\caption{Comparison of the effect of exposure to 3.7 eV light and 4.9 eV light.  Light exposure ends in both cases at time t=0 and work function measurement begins about 3 minutes later.  Pre-exposure work function data is also plotted.}\label{fig:UVact1}
\end{figure}

Exposure to 4.9 eV light has an effect on the work workfunction of the photocathode. This is shown in the bottom curve in Fig.~\ref{fig:UVact1}.  We initially measured the work function to be 2.4 eV.  The photocathode was illuminated with 4.9 eV light for 2 minutes inside the deposition chamber.  After the light exposure, the photocathode was transferred to the Kelvin Probe chamber and the work function was measured.  There is a time delay of about 3 minutes from the end of light exposure to the start of work function measurement.  A clear drop in the value of the work function by at least 150 meV was observed.  The work function appeared to recover its original value over a time period of ~30 minutes. When this experiment was repeated using a 3.7 eV light, the work function showed no obvious effect similar to that of the 4.9 eV light. (see plot in top curve of Fig.~\ref{fig:UVact1}).  It was found that 3.7 eV light produced measurable photocurrent with a QE~0.1-0.2\%, indicating that 3.7 eV is above the photoemission threshold, consistent with the literature~\cite{Sommer},\cite{PhysRevB.8.3987}.  There is a curious similarity with the result reported earlier by Sertore et al although changes in the work function were not reported.  They found that QE rejuvenation took place while simultaneously heating to $300^{\circ} $C AND illuminating with 4.9 eV light, while illuminating with 3.7 eV light produced no rejuvenation effects~\cite{Michelato1997464}.

\begin{figure}
\includegraphics[width=0.45\textwidth]{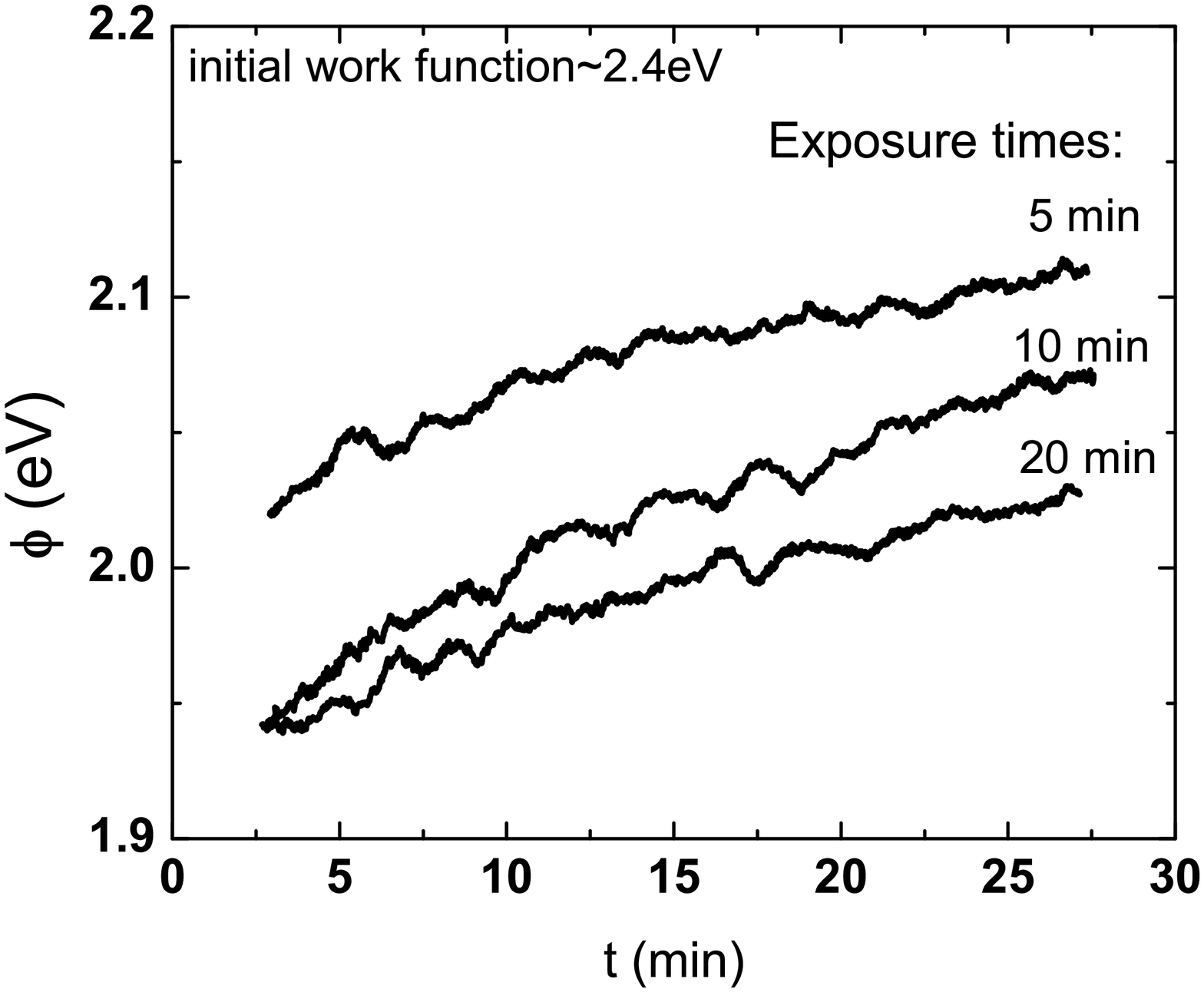}
\hfill
\includegraphics[width=0.45\textwidth]{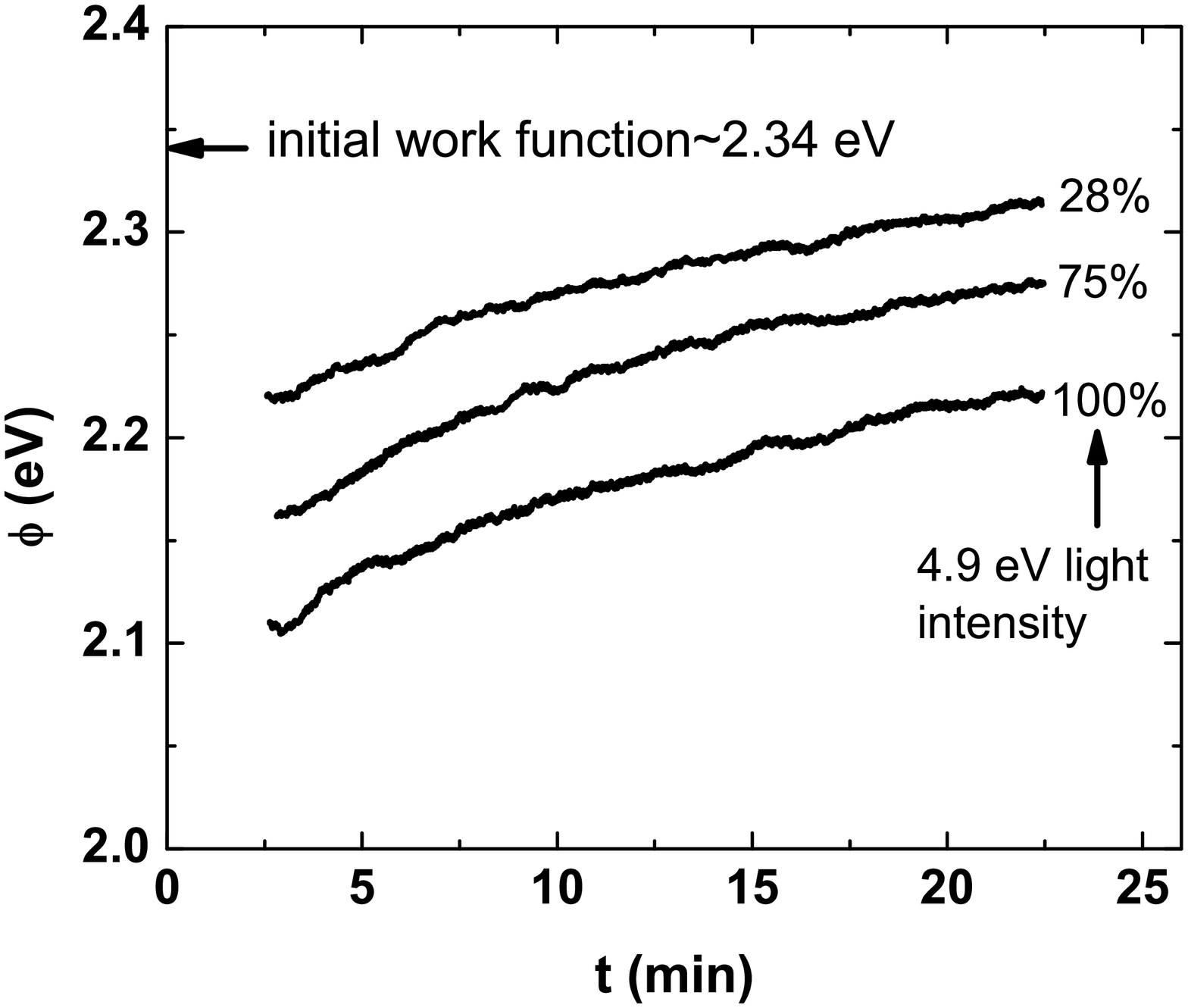}
\caption{Top:  Work function measured after various exposure times.  The initial $\Delta \phi$ saturates after about 20 minutes exposure.  Bottom: Work function measured after 2 min. exposure at different intensities.  There is still a measurable effect at 28\% intensity.}
\label{fig:Vary2}
\end{figure}

We performed two further investigations on the UV exposure effects. First, the exposure time using 4.9 eV light was varied, as shown in Fig.~\ref{fig:Vary2}, top. Longer exposure time caused a larger drop in the work function and a longer recovery time. However, it appeared that the exposure time saturates at approximately 20 minutes, whereby longer exposure time did not seem to cause the work function to drop further.

Secondly, the intensity of the 4.9 eV light was varied using neutral density filters. The cathode was illuminated for 2 minutes at a particular spot, then the cathode was moved into the Kelvin probe chamber for the work function measurement. The drop in the work function diminished as the intensity decreased. We found that the induced work function reduction scaled with the light intensity~(Fig.\ref{fig:Vary2}, bottom).

A similar observation has been reported on Kelvin probe measurements on Indium Tin Oxide (ITO)~\cite{Kim2000}.  In that work, the drop in the work function was attributed to either charging effects, or photochemistry~\cite{Kamat1993}.  We will be conducting further investigation to understand the origin of this observation in the $\mathrm{Cs}_2\mathrm{Te}$ photocathode.  As we discussed earlier, chemical changes can have significant effects on the work function.  Unfortunately to date there have been no significant studies linking changes in work function to surface chemistry.

In summary, a Kelvin Probe was used to measure the work function of $\mathrm{Cs}_2\mathrm{Te}$ photocathodes grown for the AWA drive-beam photoinjector.  The fresh cathode was found to have an initial work function of about 2.3 eV increasing to 2.8 eV as the cathode ages and the QE declines. The QE scaled inversely with the work function over time. The effect of rejuvenation via heating produced a different correlation whereby both QE and the work function increased after heating. Exposure to 4.9 eV light produced a temporary drop in the measured work function, with a recovery time on the order of 30 minutes. The magnitude of the drop in work function is dependent upon the exposure time and the intensity of the 4.9 eV light. Exposure to 3.7 eV light produced no noticeable effect.

\begin{acknowledgments}
We thank Richard Rosenberg, Wei Gai, and acknowledge valuable discussion with Karoly Nemeth. This work was funded by the U.S. Dept of Energy Office of Science under contract number DE-AC02-06CH11357 and the National Science Foundation under grant number 0969989.
The submitted manuscript has been created by UChicago Argonne, LLC, Operator of Argonne National Laboratory ("Argonne"). Argonne, a U.S. Department of Energy Office of Science laboratory, is operated under Contract No. DE-AC02-06CH11357. The U.S. Government retains for itself, and others acting on its behalf, a paid-up nonexclusive, irrevocable worldwide license in said article to reproduce, prepare derivative works, distribute copies to the public, and perform publicly and display publicly, by or on behalf of the Government.
\end{acknowledgments}

\hyphenation{Post-Script Sprin-ger}

\end{document}